
%
%
%






\newcount\refnumber
\newcount\temp
\newcount\test
\newcount\tempone
\newcount\temptwo
\newcount\tempthr
\newcount\tempfor
\newcount\tempfiv
\newcount\testone
\newcount\testtwo
\newcount\testthr
\newcount\testfor
\newcount\testfiv
\newcount\itemnumber
\newcount\totalnumber
\refnumber=0
\itemnumber=0
\def\initreference#1{\totalnumber=#1
                 \advance \totalnumber by 1
                 \loop \advance \itemnumber by 1
                       \ifnum\itemnumber<\totalnumber
                        \temp=100 \advance\temp by \itemnumber
                        \count\temp=0 \repeat}

\def\ref#1{\temp=100 \advance\temp by #1
   \ifnum\count\temp=0
    \advance\refnumber by 1  \count\temp=\refnumber \fi
   \ [\the\count\temp]}

\def\reftwo#1#2{\tempone=100 \advance\tempone by #1
   \ifnum\count\tempone=0
   \advance\refnumber by 1  \count\tempone=\refnumber \fi
   \temptwo=100 \advance\temptwo by #2
   \ifnum\count\temptwo=0
   \advance\refnumber by 1  \count\temptwo=\refnumber \fi
 \testone=\count\tempone \testtwo=\count\temptwo
 \sorttwo\testone\testtwo
     \ [\the\testone,\the\testtwo]}       

\def\refthree#1#2#3{\tempone=100 \advance\tempone by #1
   \ifnum\count\tempone=0
    \advance\refnumber by 1  \count\tempone=\refnumber \fi
    \temptwo=100 \advance\temptwo by #2
   \ifnum\count\temptwo=0
    \advance\refnumber by 1  \count\temptwo=\refnumber \fi
    \tempthr=100 \advance\tempthr by #3
   \ifnum\count\tempthr=0
    \advance\refnumber by 1  \count\tempthr=\refnumber \fi
 \testone=\count\tempone \testtwo=\count\temptwo \testthr=\count\tempthr
 \sortthree\testone\testtwo\testthr
   \test=\testthr  \advance\test by -2
 \ifnum\test=\testone    \test=\testtwo  \advance\test by -1
    \ifnum\test=\testone   
    \ [\the\testone--\the\testthr]\fi \advance\temptwo by 1
  \else
     \ [\the\testone,\the\testtwo,\the\testthr]    
 \fi}

\def\reffour#1#2#3#4{\tempone=100 \advance\tempone by #1
   \ifnum\count\tempone=0
    \advance\refnumber by 1  \count\tempone=\refnumber \fi
    \temptwo=100 \advance\temptwo by #2
   \ifnum\count\temptwo=0
    \advance\refnumber by 1  \count\temptwo=\refnumber \fi
     \tempthr=100 \advance\tempthr by #3
   \ifnum\count\tempthr=0
    \advance\refnumber by 1  \count\tempthr=\refnumber \fi
    \tempfor=100 \advance\tempfor by #4
   \ifnum\count\tempfor=0
    \advance\refnumber by 1  \count\tempfor=\refnumber \fi
 \testone=\count\tempone \testtwo=\count\temptwo \testthr=\count\tempthr
 \testfor=\count\tempfor
 \sortfour\testone\testtwo\testthr\testfor
   \test=\testthr \advance\test by -1
   \ifnum\testtwo=\test   \test=\testtwo \advance\test by -1
    \ifnum\testone=\test  \test=\testfor \advance\test by -3
     \ifnum\testone=\test \ [\the\testone--\the\testfor]
     \else \ [\the\testone--\the\testthr,\the\testfor]
     \fi
    \else  \test=\testfor \advance\test by -1
     \ifnum\testthr=\test \ [\the\testone,\the\testtwo--\the\testfor]
     \else\ [\the\testone,\the\testtwo,\the\testthr,\the\testfor]
     \fi
    \fi
   \else \ [\the\testone,\the\testtwo,\the\testthr,\the\testfor]
   \fi}

\def\reffive#1#2#3#4#5{\tempone=100 \advance\tempone by #1
   \ifnum\count\tempone=0
    \advance\refnumber by 1  \count\tempone=\refnumber \fi
    \temptwo=100 \advance\temptwo by #2
   \ifnum\count\temptwo=0
    \advance\refnumber by 1  \count\temptwo=\refnumber \fi
    \tempthr=100 \advance\tempthr by #3
   \ifnum\count\tempthr=0
    \advance\refnumber by 1  \count\tempthr=\refnumber \fi
    \tempfor=100 \advance\tempfor by #4
   \ifnum\count\tempfor=0
    \advance\refnumber by 1  \count\tempfor=\refnumber \fi
    \tempfiv=100 \advance\tempfiv by #5
   \ifnum\count\tempfiv=0
    \advance\refnumber by 1  \count\tempfiv=\refnumber \fi
 \testone=\count\tempone \testtwo=\count\temptwo \testthr=\count\tempthr
 \testfor=\count\tempfor \testfiv=\count\tempfiv
 \sortfive\testone\testtwo\testthr\testfor\testfiv
  \test=\testthr \advance\test by -1
  \ifnum\testtwo=\test   \test=\testtwo \advance\test by -1
   \ifnum\testone=\test  \test=\testfor \advance\test by -3
    \ifnum\testone=\test \test=\testfiv \advance\test by -4
     \ifnum\testone=\test\ [\the\testone--\the\testfiv]
     \else\ [\the\testone--\the\testfor,\the\testfiv]
     \fi
    \else \ [\the\testone--\the\testthr,\the\testfor,\the\testfiv]
    \fi
   \else  \test=\testfor \advance\test by -1
    \ifnum\testthr=\test \test=\testfiv \advance\test by -2
     \ifnum\testthr=\test \ [\the\testone,\the\testtwo--\the\testfiv]
     \else \ [\the\testone,\the\testtwo--\the\testfor,\the\testfiv]
     \fi
    \else\ [\the\testone,\the\testtwo,\the\testthr,\the\testfor,\the\testfiv]
    \fi
   \fi
  \else \test=\testfor \advance\test by -1
   \ifnum\testthr=\test \test=\testfiv \advance\test by -2
    \ifnum\testthr=\test\
[\the\testone,\the\testtwo,\the\testthr--\the\testfiv]
    \else\ [\the\testone,\the\testtwo,\the\testthr,\the\testfor,\the\testfiv]
    \fi
   \else\ [\the\testone,\the\testtwo,\the\testthr,\the\testfor,\the\testfiv]
   \fi
  \fi}

\def\refitem#1#2{\temp=#1 \advance \temp by 100 \setbox\count\temp=\hbox{#2}}

\def\sortfive#1#2#3#4#5{\sortfour#1#2#3#4\relax
   \ifnum#5<#4\relax \test=#5\relax #5=#4\relax
     \ifnum\test<#3\relax #4=#3\relax
       \ifnum\test<#2\relax #3=#2\relax
         \ifnum\test<#1\relax  #2=#1\relax  #1=\test
         \else #2=\test \fi
       \else #3=\test \fi
     \else #4=\test \fi \fi}

\def\sortfour#1#2#3#4{\sortthree#1#2#3\relax
    \ifnum#4<#3\relax \test=#4\relax #4=#3\relax
       \ifnum\test<#2\relax #3=#2\relax
          \ifnum\test<#1\relax #2=#1\relax #1=\test
          \else #2=\test \fi
       \else #3=\test \fi \fi}

\def\sortthree#1#2#3{\sorttwo#1#2\relax
       \ifnum#3<#2\relax \test=#3\relax #3=#2\relax
          \ifnum\test<#1\relax #2=#1\relax #1=\test
          \else #2=\test \fi \fi}

\def\sorttwo#1#2{\ifnum#2<#1\relax \test=#2\relax #2=#1\relax #1=\test \fi}


\def\setref#1{\temp=100 \advance\temp by #1
   \ifnum\count\temp=0
    \advance\refnumber by 1  \count\temp=\refnumber \fi}

\def\printreference{\totalnumber=\refnumber
           \advance\totalnumber by 1
           \itemnumber=0
           \loop \advance\itemnumber by 1  
                 \ifnum\itemnumber<\totalnumber
                 \item{[\the\itemnumber]} \unhbox\itemnumber \repeat}
%
%
%
  \magnification=1200
  \hsize=15.5 truecm
  \vsize=23.0truecm
  \topskip=20pt            
  \fontdimen1\tenrm=0.0pt  
  \fontdimen2\tenrm=4.0pt  
  \fontdimen3\tenrm=7.0pt  
  \fontdimen4\tenrm=1.6pt  
  \fontdimen5\tenrm=4.3pt  
  \fontdimen6\tenrm=10.0pt 
  \fontdimen7\tenrm=2.0pt  
  \baselineskip=17.0pt plus 1.0pt minus 0.5pt  
  \lineskip=1pt plus 0pt minus 0pt             
  \lineskiplimit=1pt                           
  \parskip=2.5pt plus 5.0pt minus 0.5pt
  \parindent=15.0pt
%
%
\font\sub=cmbx10 scaled\magstep1
 2
\font\bx=cmr8
\def\lsim{\; \raise0.3ex\hbox{$<$\kern-0.75em\raise-1.1ex\hbox{$\sim$}}\; }
\def\gsim{\; \raise0.3ex\hbox{$>$\kern-0.75em\raise-1.1ex\hbox{$\sim$}}\; }
\def\jump{\vskip 1truecm}
\def\GeV{\rm GeV}
\def\MeV{\rm MeV}

\def\fig#1{{\bf Figure #1}.}

\def\cO{{\cal O}}

\def\1{\aa}

\def\del{\partial}

%
%

%

\def\mpl#1#2#3{{\it  Mod.\ Phys.\ Lett.\ }{{\bf #1} {(#2)} {#3}}}

\def\pr#1#2#3{{\it Phys.\ Rev.\ }{{\bf #1} {(#2)} {#3}}}

\def\pl#1#2#3{{\it  Phys.\ Lett.\ }{{\bf #1} {(#2)} {#3}}}
\def\prep#1#2#3{{\it Phys.\ Rep.\ }{{\bf #1} {(#2)} {#3}}}

\def\rmp#1#2#3{{\it  Rev.\ Mod.\ Phys.}{{ \bf #1} {(#2)} {#3}}}

\def\zp#1#2#3{{\it Z. Phys.} {{\bf #1} {(#2)} {#3}}}
\def\ibid#1#2#3{{\sl ibid.\ }{{\bf #1} {(#2)} {#3}}}

\def\frac#1#2{{#1\over#2}}
\def\ann#1#2{#1\leftrightarrow #2}
\def\Gann{\Gamma_{\rm ann}}
%
%
\newcount\eqnumber
\eqnumber=1
\def\chaphead{}

\def\new{\hbox{(\chaphead\the\eqnumber}\global\advance\eqnumber by 1}
\def\eqref#1{\advance\eqnumber by -#1 (\chaphead\the\eqnumber
     \advance\eqnumber by #1 }
\def\first{\hbox{(\chaphead\the\eqnumber{a}}\global\advance\eqnumber by 1}
\def\last{\advance\eqnumber by -1 \hbox{(\chaphead\the\eqnumber}\advance
     \eqnumber by 1}
\def\eq#1{\advance\eqnumber by -#1 equation (\chaphead\the\eqnumber
     \advance\eqnumber by #1}
\def\eqnam#1{\xdef#1{\chaphead\the\eqnumber}}


\def\eqt#1{Eq.~({{#1}})}

\def\gm{g(\mu)}
\def\qcd{T_{\rm QCD}}
\def\qq{q\bar{q}}
\def\lqcd{\Lambda_{\rm QCD}}
\def\Gann{\Gamma_{\rm ann}}
\def\dlips#1#2{\frac{d^3#1_{#2}}{(2\pi)^3E_{#2}}}
\def\lopo#1#2{\ln^{#2}\left(\frac s{m_{#1}^2}\right)}
\def\Fg{F^{2g}}
\def\Fgg{F^{3g}}

In the early universe chemical equilibrium between particles like gluons
and quarks was sustained by annihilations. On dimensional grounds, the
$\ann 22$ annihilation cross sections at energies much higher than the
masses of the particles involved should decrease as
$g^4(\mu)s^{-1}\sim g^4(T)T^{-2}$, where $\gm$ is the coupling
renormalized at the scale $\mu$. Thus at very high temperatures, and in
a radiation dominated universe, the total averaged annihilation rate
should scale as $\Gann^{\rm tot}\sim g^4(T)T$.  One should then compare
$\Gann^{\rm tot}$ with the Hubble rate $H=1/2t\sim T^2$, and it is
rather obvious that at sufficiently high temperatures, and for
sufficiently weak interactions, $\ann 22$ annihilations could not have
maintained equilibrium. For instance, given  only the QCD interactions
which become, on the average\ref{1}, weaker as $T\to\infty$, at some
high temperature quarks and gluons may not have been in thermal contact
at all. If so, it certainly would have interesting ramifications for our
view of the very early universe as ideal gas in thermal equilibrium. In
fact, it was already roughly estimated in\ref{2} that in SU(5) grand
unified theory equilibration is possible only at temperatures $T\lsim
3\times 10^{15}\GeV$, and if the initial state is far out of
equilibrium, chemical equilibrium is established at
temperatures much lower than this\ref{3}. Whether a grand unified
theory actually exists or not is of
course unknown, but it is very likely that QCD interactions were in full
thermal equilibrium prior to the QCD phase transition at $T=\qcd\simeq
200\ \MeV$. Given that, an interesting question then is, within the
framework of just the Standard Model, at which temperature did chemical
equilibrium between quarks and gluons first become possible?

In the present paper we address this question by considering carefully
$\qq$ annihilation into gluons in the early universe. The lowest order
process is $\qq\to gg$, but one expects that when $s\gg \lqcd^2$, $\qq$
annihilation into many gluons becomes more and more important. Therefore
in the early universe it might be essential to consider also $\qq\to
gg\cdots g$--annihilations. We show however that at very high
temperatures these modify  the annihilation rate by only about 3\%.
Comparing the total annihilation rate with the expansion rate of the
universe, which we correct for the Standard Model interactions in the
primeval plasma, we find that quarks and gluons were not in chemical
contact above $T\gsim 3\times 10^{14}~\GeV$.  We expect that at still
higher temperatures the Standard Model particles did not have any
thermal contact whatsoever, but precise study of kinetic equilibrium
between e.g.~gluons is  made difficult by infrared problems and soft
gluon physics, which cannot be dealt with succesfully in perturbation
theory, and we do not address this issue here.

Let us begin by computing the thermally averaged annihilation rate
$\Gann (\qq\to gg)$, which in the limit when the final state blocking
is neglected can be defined as\eqnam{\average}
$$
\Gann (\qq\to gg)=\frac 1{n_q}\int \dlips{p}{1}\dlips{p}{2}
f(E_1/T)f(E_2/T)\sigma (\qq\to gg) v_{rel}p_1\cdot p_2 \eqno\new)
$$
where $n_q$ is the quark density (and equal to the anti--quark density),
$f(E/T)=(\exp(E/T)+1)^{-1}$ and $v_{rel}=\sqrt{1-\left(m^2/ (p_1\cdot
p_2)\right)^2}$ is the invariant velocity. The relevant $t$-- and
$s$--channel diagrams are depicted in Fig.\ 1. We find that \eqnam{\22x}
$$
\eqalign{&\cr
\sigma (\qq\to gg)= \frac{2\pi\alpha_s^2}{N^2s} & \left[\left(
\frac{2x^2+2x-1}{x(x-1)}  \ln(\sqrt{x}+\sqrt{x-1})
-\frac{x+1}{\sqrt{x(x-1)}}\right)\right.B\cr
  &+ \left.\left(\frac 1{2x(x-1)}\ln(\sqrt{x}+\sqrt{x-1})-\frac
1{12}\frac{4x+5}
{\sqrt{x(x-1)}}\right)A\right], \cr &\cr }
\eqno\new)
$$
where $x=s/4m^2$ with $m$ the quark mass, and
$A=C_AT_F(N^2-1)=N/2(N^2-1)$ and $B=NC_F^2=1/(4N)(N^2-1)^2$ are $SU(N)$
color factors. Note that in the case of QED, $A=0$ and $B=1$ so that
\eqt{\22x} reduces to the well known Dirac formula.

In the early universe it is natural to replace $m$ by the quark plasma
mass $m_q(T)$ in \eqt{\average}. At high temperature the left--handed
quark and gluon plasma masses are given by  \ref{7}\eqnam{\plasma}
$$
\eqalign{
m^2_g(T)=&\frac 23g^2_sT^2,\cr
m^2_q(T)=&(\frac 16g^2_s+\frac 3{32}g^2_W+\frac 1{288}g^2_Y)T^2,\cr}
\eqno\new)
$$
where $g_W$ is the weak and $g_Y$ the U(1)$_Y$ coupling; for
right--handed quarks the purely weak contribution  is absent.
Substituting these back to Eqs.\ (\average) and (\22x) we find that
$\Gann (\qq\to gg)$ can be written as \eqnam{\22ave}
$$
\Gann (\qq\to gg)={\alpha^2_s(T)\over 288\pi\zeta(3)}\Fg (\log\; T)T,
\eqno\new)
$$
where $\Fg (\log\; T)$ is dimensionless quantity that can be solved
numerically. Eq.\ (\plasma) implies that $\Fg$ can depend on $T$ only
through running of  $\alpha_s$ with temperature. Thus  it is a smooth,
slowly changing function  of $T$ and, neglecting $\rm SU(2)\times U(1)$
corrections in the quark plasma masses,   $\Fg_0(3)= 431$ and
$\Fg_0(15)= 680$ (see Table 1). If all quarks were left--handed, we
would obtain $\Fg_1(3)= 397$ and $\Fg_1(15)= 597$. In what
follows, we do not compute the rate (\22ave) separately for different
chiralities but  approximate $\Fg$ by the average of $\Fg_0$ and
$\Fg_1$.

We should comment here that the QCD running coupling at finite
temperature is, in general, a problematic concept, and there are several
conflicting statements about the way it scales with
temperature\reftwo{8}{11}. This has to do with the fact that at finite
temperature, the system has two a priori independent mass scales, $T$
and the renormalization point $\mu$, so that the limit $T\to\infty$ is
not unambigious. In our case thermal averaging, \eqt{\average}, tacitly
assumes that the external legs are, apart from plasma mass corrections,
free particles with momenta partitioned thermally. (This should be a
good approximation for weakly interacting gas.) In that case the
ensemble average $\bar{G}$ of the effective charge\ref{1}, defined as
the thermal average of the usual effective charge, can be shown to scale
as $\bar{G}(\mu,T,g)\to\bar{G}(\mu,T,\bar{g})$ under $T\to\lambda T$,
where $\bar{g}=\bar{g}(\lambda\mu,T/\mu)$ is the running coupling with
$T/\mu$ fixed. Thus, in collisions the coupling is, on the average, the
running finite $T$ coupling\ref{8}\eqnam{\coupling}
$$
g^{-2}_s(\mu,T)={g^{-2}_s(\mu_0,0)}{+\frac{1}{16\pi^2}\left [7\ln
\left({{\mu}\over {\mu_0}}\right)^2+a_0(T/\mu)-a_0(0)\right]
}
\eqno\new)
$$
with $T/\mu$ fixed. The origin of the function $a_0$, which in the
$T/\mu\to\infty$ limit is a polynomial of $T/\mu$, are both the vacuum
and the $T$--dependent finite parts appearing in the charge
renormalization constant. That is, $a_0(T/\mu)$ gives the first order
change to the running coupling $g_s$ when changing the renormalization
scheme from the $MS$--scheme: \eqnam{\gdiff}
$$
g=g_{MS}\left(1+a_0(T/\mu)g_{MS}^2+\cO (g_{MS}^4)\right)
\eqno\new)
$$
Thus minimal sensitivity to the temperature corrections is obtained
when $a_0$ vanishes. In the $MOM$--scheme for pure QCD this happens
when  $\mu\simeq 2.6\; T$\ref{1}. With this particular $T/\mu$ value
$g_{MOM}$ runs exactly like $g_{MS}(T=0)$:
\eqnam{\gmom}
$$
g^{-2}_{MOM}(\mu,T)={g^{-2}_{MS}(\mu_0,0)}{+\frac{11}{16\pi^2}\ln
\left({{\mu}\over {\mu_0}}\right)
}
\eqno\new)
$$
In what follows we shall assume that one may also here adopt a prescription
where $a_0(T/\mu)=0$. Then we are able to convert the value of the
strong coupling measured in LEP at $\mu\simeq M_{\rm Z}$ and $T=0$
\ref{9}, $\alpha_s=0.12$, to finite temperatures by using
\eqt{\coupling}. In effect this means that $g_s$ (as well as $g_W$ and
$g_Y$) runs exactly as in vacuum, but with $\mu$ replaced by $\kappa T$
with $\kappa$ a constant, and the LEP reference point is translated to
the temperature $T\simeq M_{\rm Z}/\kappa$.

The computation of $\qq\to ggg$ is more involved as compared with
$\qq\to gg$ because of the more complicated kinematics of $\ann 23$
scattering. The relevant diagrams and definitions of kinematical
variables are shown in Fig.\ 2.  We perform the calculation in  large--N
approximation where the leading part of the the matrix element for
massless quarks and gluons, summed over colour and helicity degrees of
freedom, can be written  as\ref{6} \eqnam{\matrix}
$$
|M(\qq\to ggg)|^2=2g_s^{6}N^{2}(N^2-1)\frac 1s \sum^n_{i=1}\left(
s^3_{qi}s_{\bar{q}i} +s_{qi}s^3_{\bar{q}i}\right) \sum_{1,2,3}{1\over
s_{q1}s_{12}s_{23}s_{3\bar q}}+\cO (N^{-2}) \eqno\new)
$$
where $s_{ij}=(p_i+p_j)^2$,  $s_{qi}=(p_q-p_i)^2$ and $s_{\bar q
i}=(p_{\bar q}-p_i)^2$. The second sum in \eqt{\matrix} extends over
permutations of the gluon indices.

The $\ann 23$ cross section can be written, in the notation of\ref{10},
as\eqnam{\23x}
$$
\sigma (\qq\to ggg)={\alpha_s^3(N^2-1)\over 32\pi
s\lambda(s,m_q^2(T),m^2_q(T))}\int
{dt_1dt_2ds_1ds_2\over (-\Delta_4)^{\frac 12}}{P_4(s_1,s_2,t_1,t_2) \over
s_1s_2t_1t_2},
\eqno\new)
$$
where $\Delta_4$ is the Gram determinant defined in\ref{10} and $s_i$
and $t_i$ are invariants defined in Fig.\ 2. The term $P_4$ is  the
first sum in \eqt{\matrix} with invariants $s_{ij}$ written in  terms of
$s_i$ and $t_i$ in the center--of--mass frame.  Although the cross
section \eqt{\23x} has infrared problems, in the early universe the
plasma masses of quarks and gluons act as natural  regulators. We
account for them by  replacing  $s_i, t_i\to s_i-m^2,t_i-m^2$ in
\eqt{\23x}, where $m$ is the relevant plasma mass.  Note that plasmon
decay into quarks is not kinematically possible so that $\ann 12$
processes should not play a major role in equilibration.

We have evaluated the cross section (\23x) numerically, but when $s\gg
m^2$, we find that it is well fitted by\eqnam{\fit}
$$
\eqalign{
\sigma & (\qq\to ggg)\simeq {\alpha_s^3\over s}\Bigl[ -251+28.7\lopo{q}{}
+22.1\lopo{g}{}-1.01\lopo{q}{2}\cr
&-4.33\lopo{q}{}\lopo{g}{}+2.46\lopo{g}{2}+
0.67\lopo{q}{}\lopo{g}{2} \Bigr]\cr}
\eqno\new)
$$
The fit is valid when $s/m^2\gsim 10^3$. Comparing with $\sigma (\qq\to
gg)$ we find that annihilation into three gluons, taking $m_g=2m_q$, is
equally important when $s/m_q^2\simeq 10^4$ with  $\alpha_s\simeq 0.12$.
For very large $s/m_q^2$, 3--gluon final states begin slowly to dominate
over 2--gluon final states. For example, when $s/m_q^2\simeq 10^{11},\
\sigma(\qq\to ggg)/\sigma(\qq\to gg)\simeq 5$.

In the early universe $s$ in collisions was on the average about
$18T^2$, not significantly larger than the squared thermal masses.
Therefore one should not expect an enhancement of the average
annihilation rate due to the large logarithms of the $\qq\to ggg$ cross
section. We have evaluated  the thermally averaged $\qq\to ggg$ rate as
in \eqt{\average}, and we find numerically that\eqnam{\23ave}
$$
\Gann  (\qq\to ggg)= {1\over 2\pi^2\zeta(3)}\Fgg (\log T)\alpha^3_s(T)T
\eqno\new)
$$
where $\Fgg$ is tabulated in Table 1, and we see that at very high
temperatures $\ann 23$ rate is  about 3\%\ of the $\ann 22$ rate. Thus,
although quarks and antiquarks annihilate into any number of gluons in
the early universe, these processes should not contribute
significantly to the total annihilation rate $\Gann^{\rm tot}$.

We now compare the total averaged annihilation rate $\Gann^{\rm tot}$
with the expansion rate of the universe as given by the Hubble parameter
$H=\sqrt{8\pi G_N\rho/3}$. In what follows we shall take into account
the fact that the gas in the early universe is interacting, which
modifies  the energy density $\rho$ from its ideal gas value. The
thermodynamic potential $\Omega$ has been computed for SU(N) gauge
theories in perturbation theory\ref{5}, and to lowest it reads
\eqnam{\exc}
$$
\eqalign{&\cr
\Omega_0=& -{\pi^2T^4\over 45}\left( N^2-1+\frac 74 NN_f\right),\cr
\Omega_1=& {(N^2-1)g^2T^4\over 144}\left( N+\frac 54 N_f\right),\cr &\cr}
\eqno\new)
$$
where $\Omega_0$ is thermodynamic potential for ideal gas and $\Omega_1$
is the lowest order exchange energy correction, and $N_f$ is the number
of  flavours. For U(1) one should set $N^2-1\to 0,~N\to 1$ in $\Omega_0$
and $N^2-1\to 1,~N=0,~N_f\to\sum Q_f^2$ in $\Omega_1$.  Energy density
is given by $\rho=\Omega+TS=\Omega-T\del\Omega/\del T=-3\Omega$ with
$\Omega=\Omega_0+\Omega_1$. In the Standard Model at $T\gg 100\ \GeV$ we
find that\eqnam{\rhosm}
$$\rho_{\rm SM}={\pi^2T^4\over
30}\left(g_*(T)-{5\over 2\pi}\Bigl(
84\alpha_s+\frac{57}2\alpha_W+\frac{25}{12}\alpha_Y\Bigr)\right),
\eqno\new)
$$
so that at high temperature $\rho_{\rm SM}$ differs from  ideal gas
value by a few percents. The effect of the exchange energy  in the case
of the Standard Model is illustrated  in Fig.\ 3 for $T\gg 100$ GeV,
assuming Higgs and top masses can be neglected.

In Fig.\ 4 we have drawn $H$ together with $\Gann^{\rm tot}\simeq\Gann
(\qq\to gg)+\Gann (\qq\to ggg)$. We see that the total annihilation rate
is too slow to maintain chemical equilibrium when\eqnam{\final}
$$
T=T_d\gsim 2.5\times 10^{14} ~\GeV.
\eqno\new)
$$

The effects considered here work all in the same direction: including
$\qq\to ggg$ in $\Gann^{\rm tot}$ increases the decoupling temperature
$T_d$, and so does also the exchange energy (\exc) by slowing down the
Hubble rate. Annihilation into four or more gluons, which in principle
could increase $T_d$ further, is unlikely to change the conclusion
(\final) because of the smallness of the logarithmic enhancement of the
multiparticle production rate. Kinetic equilibrium, which is maintained
by elastic collisions, may extend to temperatures somewhat higher than
$T_d$, but also kinetic contact is lost at sufficiently high
temperature. Thus,  unless the quark and gluon ensembles were, for some
reason, created thermal at energy scales $M_{Pl}$, they cannot have had
thermal distributions at and above GUT scales. The same conclusion would
naturally hold also for the rest of the Standard Model particles. This
is directly relevant to the treatment of phase transitions at very high
temperatures, which often is based on the assumption that the background
Higgs field (or the scalar order parameter) evolves in a thermal
background. In reality the familiar thermal corrections to the effective
potential above $T_d$ may be completely absent, or present only in a
modified form.
\jump
\centerline{\sub Acknowledgements}\jump
We thank Paul Hoyer for useful information on scattering in QCD.

\vfill\eject
\centerline{\sub References}
\jump
\refitem{1}{K.\ Enqvist and K. Kainulainen, \zp {\bf C53}{1992}{87}.}
\refitem{2}{J.\ Ellis and G.\ Steigman, \pl{B89}{1979}{186}.}
\refitem{3}{K.\ Enqvist and K.J.\ Eskola, \mpl{A5}{1990}{1919}.}
\refitem{4}{LEP}
\refitem{5}{See T.\ Toimela, {\it Int.\ J.\ Theor.\ Phys.\ }{\bf 24} (1985) 901
and references therein; {\it erratum} {\bf 26} (1987) 1021.}
\refitem{6}{M.L.\ Mangano and S.J.\ Parke, \prep{C200}{1991}{186}.}
\refitem{7}{H.A.\ Weldon, \pr{D26}{1982}{1394}; \ibid {D26}{1982}{2789}; D.J.\
Gross, R.D.\ Pisarski and L.G.\ Yaffe, \rmp{53}{1981}{43}.}
\refitem{8}{Y.\ Fujimoto and H.\ Yamada, \pl{B195}{1987}{231}; {\sl ibid.\
}{\bf B200} (1988) 167.}
\refitem{9}{T.\ Hebber, \prep{C317}{1992}{69}.}
\refitem{10}{E.\ Byckling and K.\ Kajantie, {\it Particle Kinematics}, John
Wiley and Sons,
(London, New York, Sydney, Toronto, 1973).}
\refitem{11}{H.\ Matsumoto, Y.\ Nakano and H.\ Umezawa, \pr{D29}{1984}{1116};
K.\ Enqvist and K.\ Kajantie, \mpl{A2}{1987}{479}; H.\ Naggawa, A.\ Niegawa and
H.\ Yokota, \pr{D38}{1988}{2566}; R.\ Baier, B.\ Pire and D.\ Schiff,
\pl{B238}{1990}{367}.}
\printreference
\def\jumpp{\jump\noindent}
\vfill\eject\null\jumpp
\centerline{\sub Table caption}\jumpp
\def\xrule{\vrule height1pt width7truecm depth0pt\par}
\def\xruler{\vrule height4pt width7truecm depth-3pt\par}
\def\hhh{&&\hskip 5pt}
 \settabs 8\columns{\+ &&\xruler  & & & & & \cr
   \+ &&$\log\ T$ & $F^{2g}_0$ & $F^{2g}_1$ & $F^{3g}$ \cr
\+ &&\xrule &&& \cr
\+  \hhh 3    & 431  & 397 & 3  &&\cr
\+ \hhh  6    & 525  & 474 & 7 &&\cr
\+ \hhh 9    & 590  & 526 & 10 &&\cr
\+ \hhh 12   & 640  & 565 & 15 &&\cr
\+ \hhh 15   & 680  & 597 & 19 &&\cr
\+ &&\xrule & & && &\cr }
\jumpp
Table 1. The behaviour of the functions $F^{2g}_0$, $F^{2g}_1$ and
$F^{3g}$ with temperature. $F^{2g}_0$ contains only the $SU(3)_c$ contribution
to the thermal particle masses, while $F^{2g}_1$ accounts also for the
electroweak corrections but assumes that all quarks are left--handed.
$F^{3g}_0$  has only $SU(3)_c$ corrections.
\vfill\eject\null\jumpp
\centerline{\sub Figure captions}
\jumpp
\fig 1 The annihilation diagrams for $\qq\to gg$.\jumpp
\fig 2 The annihilation diagrams for $\qq\to ggg$, and the
definitions of the relevant kinematical variables.\jumpp
\fig 3 The Standard Model degrees of freedom at high temperatures, corrected
for the lowest order
exchange energy (solid curve). For comparision, we show also the ideal gas
value $g_*=106.75$ (dashed line).\jumpp
\fig 4 Comparision of the Hubble rate (solid curve), as computed from (\rhosm),
with the
$\qq$ annhilation rate $\Gann^{\rm tot}$ (dashed line).
\vfill\eject
\hoffset=0pt\null
\nopagenumbers
\vskip -20pt
\line {\hfill NORDITA--93/32 P}
\vskip 1.2truecm
\centerline {\sub Chemical equilibrium in QCD gas in the early universe }
\vskip 1.5truecm
\centerline {  Kari Enqvist{\footnote{\bx
$^\dagger$}{\baselineskip=9pt\bx
internet: enqvist@nbivax.nbi.dk}}}
\vskip 0.4truecm
\centerline {Nordita, Blegdamsvej 17, DK-2100 Copenhagen \O , Denmark}
\vskip 0.4truecm
\centerline{and}\vskip 0.4truecm
\centerline{Jukka Sirkka{\footnote{\bx $^\star$}{\baselineskip=9pt\bx
internet: sirkka@sara.cc.utu.fi }}}
\vskip 0.4truecm
\centerline {Department of Physics, University of Turku, SF-20500 Turku,
Finland}
\jump
\centerline {\bf Abstract}
\jump
\noindent
We compute the thermally averaged $\qq$--annihilation rate into two and
three gluons in the early universe. We show that at very high
temperatures $\qq\to ggg$ represents only a 3\%\ correction
to $\qq\to gg$. Comparing the annihilation rate to the Hubble rate,
corrected for particle interactions in the Standard Model gas, we
find that quarks and gluons are not in chemical equilibrium when $T\gsim
3\times 10^{14}$ GeV.
\vfill\eject
\end